\documentclass[preprint,12pt]{elsarticle}

\usepackage{amssymb}
\usepackage{amsmath}
\journal{Acta Astronautica}

\begin{document}

\begin{frontmatter}



\title{A search for Planet Nine with small spacecraft: Three-body, post-Newtonian, non-gravitational, planetary and Kuiper Belt effects}


\author[1] {Sahin Ulas Koprucu \corref{cor1}}
\ead{ulas.koprucu@tubitak.gov.tr}
\cortext[cor1]{Corresponding author}
\affiliation[1]{organization={TUBITAK UZAY - Space Technologies Research Institute},%Department and Organization
            city={Ankara},
            country={Turkey}}

\author[2] {Bayram Tekin}
\ead{btekin@metu.edu.tr}
\affiliation[2]{organization={Department of Physics, Middle East Technical University},%Department and Organization
            city={Ankara},
            country={Turkey}}

\begin{abstract}

A hypothetical gravitating body in the outer Solar System, the so-called Planet Nine, was proposed to explain the unexpected clustering of the Kuiper Belt Objects. As it has not been observed via telescopes, it was conjectured to be a primordial black hole (of the size of a quince) that could be gravitationally detected by laser-launching or solar sailing many small spacecraft. Here, we study various aspects that will affect such a search for Planet Nine. Our basic observable is the angular displacement in the trajectory of a small spacecraft which will be mainly affected by the gravity of Planet Nine, augmented with several other 3-body, non-gravitational, post-Newtonian, planetary, and Kuiper Belt effects. First, we calculate the effect of the Sun in the framework of the circular restricted three-body problem of the Sun--Planet Nine-spacecraft for the two particular initial conditions. Then, we study the effects of Kuiper Belt and outer planets, namely Jupiter, Saturn, Uranus, Neptune, as well as non-gravitational perturbations such as magnetic and drag forces exerted by the interstellar medium; and the solar radiation pressure. In addition, we investigate the post-Newtonian general relativistic effects such as the frame-dragging, Schwarzschild effect, and geodetic precession on the spacecraft trajectory. We show that the leading order angular displacement is due to the solar radiation pressure for the lower spacecraft velocities, and the drag force for the higher spacecraft velocities. Among the general relativistic effects, the frame-dragging has the smallest effect; and the Schwarzschild effect due to Sun has the largest effect. However, none of the general relativistic effects produces a meaningful contribution to the detection.

\end{abstract}

\end{frontmatter}


\section{Introduction} 

The possible existence of a body at a distance 500 AU, with a mass around 5--10 $M_{\oplus}$,  commonly known as Planet Nine, was proposed to explain the unexpected clustering of the Kuiper Belt Objects (KBO) ~\citep{BB1,Batygin2019}. The earlier orbital constraints for a distant Solar System body from the planetary data was investigated in~\citep{Iorio1}. To provide additional evidence, the generation of highly inclined and also low-inclination, Neptune-crossing Trans-Neptunian Objects due to Planet Nine were studied in~\citep{BBnew1,BBnew2}. In addition, the effect of Planet Nine on the obliquity of Sun and Uranus were also investigated in~\citep{BBnew3, Lu}. As an alternative explanation of the clustering of the KBO, modified gravity theory was also proposed instead of invoking a planet ~\citep{mond}. So far, detection has not yet been achieved, telescope searches are still ongoing and there is a lot of room in this volume of space for such a planet to hide~\citep{atacama}. But in the absence of visual observation, an exciting conjecture was put forward: this object could be a primordial black hole ~\citep{blackhole}. In that case, its size would be around the size of a medium-sized quince and alternative ways for detection have been suggested such as sending a cluster of small sub-relativistic spacecraft and measuring the deflection of their trajectories due to the gravity of Planet Nine~\citep{witten}; or aiming to measure the Hawking radiation spreading from the primordial black hole with a sub-relativistic spacecraft, although capturing the weak signal would be challenging ~\citep{hawking}. In addition, the Zwicky Transient Facility public archive shows no candidate~\citep{BB2}. The use of radio tracking measurements of New Horizons spacecraft was proposed to put constraints on the location of a distant hypothesized Solar System body~\citep{Iorio2}. The similar idea was used to investigate existence of Mars-sized perturber at 65--80~AU by using the Cassini spacecraft measurements~\citep{Iorio3}. Then, the use of accurate trajectory of Saturn from the ranging data of Cassini spacecraft to put constraint on the location of Planet Nine was investigated~\citep{Iorio4,Fienga} and the name Telisto, meaning farthest, was also suggested instead of Planet Nine~\citep{Iorio4}. More recently, an indirect way was suggested in~\citep{chan} assuming that Planet Nine has satellites, the tidal force of Planet Nine can heat up the putative satellites causing thermal radio flux for observations. In another recent work, a candidate location of Planet Nine was proposed by tracing the trajectory of an interstellar meteoroid backward and looking at its coincidence with the maximum probability region of Planet Nine ~\citep{navarro}. Lastly, the localization of Planet Nine was suggested in~\citep{bucko} by analyzing the trajectory of a spacecraft that will be sent for a possible Uranus mission. 

In this work, our goal is to study all possible perturbations on the trajectory of a small spacecraft which will be sent to detect Planet Nine and investigate whether the effect of Planet Nine can be distinguishable from the other perturbations. We focus on the method proposed in~\citep{witten} based on probing the gravitational field of Planet Nine with a spacecraft attaining sub-relativistic speed launched with the use of advanced propulsion technologies such as laser propulsion~\citep{starshot} but particularly solar sail~\citep{solarsail, solarsail2} which is partially tested technology. The sub-relativistic speed was chosen to make the mission duration in a reasonable time interval, like order of decades. In~\citep{witten}, the deflection was studied in terms of time delay in the signal sent from the spacecraft to the Earth with sufficiently accurate timekeeping requirements. To avoid this requirement, measurement of the transverse displacement was proposed by using Very Long Baseline Interferometry (VLBI)~\citep{bruteforce}. In addition, the effects of the drag and electromagnetic forces exerted by the interstellar medium on the spacecraft’s trajectory were also discussed in~\citep{loeb1}, including the parameter space of the spacecraft to distinguish the gravity of Planet Nine from other perturbations.

In~\citep{witten}, the deflection was studied in the context of Planet Nine and the spacecraft as a two-body problem. We extended that discussion by including the Newtonian gravity of the Sun; and studied the deflection in the circular restricted three-body problem (CRTBP) of the Sun--Planet Nine-spacecraft with two particular initial conditions under some assumptions. Then, the effects of Kuiper Belt and outer planets, namely Jupiter, Saturn, Uranus, Neptune, as well as the non-gravitational perturbations such as drag force, magnetic force, and solar radiation pressure were investigated to assess whether the effect of Planet Nine can be distinguishable from the other perturbations. In fact, the effects of drag and magnetic forces were already studied in~\citep{loeb1} in terms of magnitude comparison. Here, we performed the numerical integration by considering the directions as well, and the nominal trajectory of the spacecraft is defined with the Sun and the spacecraft's two-body problem. As distinct from~\citep{loeb1}, we also discuss the general relativistic effects like the frame-dragging effect, the Schwarzschild effect, and the geodetic precession effect on the spacecraft trajectory and give the magnitude comparison with the Newtonian gravity of Planet Nine. 

The layout of this paper is as follows: Section II is a recap of~\citep{witten} where we give more details on Planet Nine and the spacecraft two-body problem as a preparation for Section III where we study the effects of the Sun within the context of CRTBP. In section IV we study the planetary and Kuiper Belt effects; in section V we study the drag force on the spacecraft trajectory; in section VI we study the magnetic force; in section VII we study the solar radiation pressure while section VIII is devoted to the post-Newtonian general relativistic effects. As the issue here is to detect a quince-sized object in such a large volume of space, we wanted to make sure that we take all the known effects into consideration and compute their contributions to the trajectory of the tiny spacecraft. Qualitatively, one knows before the computations that these effects will be small, but our aim here is to carefully quantify the effects.  
\section{Planet Nine and Spacecraft Two-Body Problem} 
\begin{figure}[h]
\centering
\includegraphics[width=.3\textwidth]{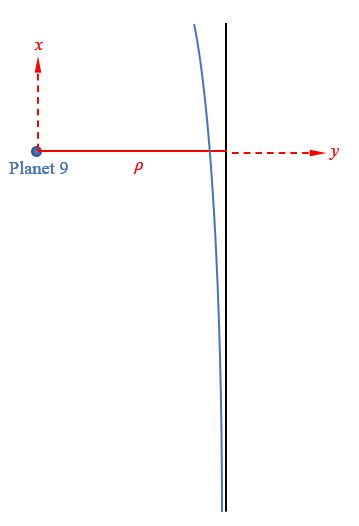}
\caption{The black line is the unperturbed trajectory and the blue curve is the perturbed trajectory due to Planet Nine; also $\rho$ is the impact factor i.e. the minimum distance to the Planet Nine along the unperturbed trajectory.}
\label{fig:witten}
\end{figure}

As a way to prepare for the three-body problem of the next section, and define the parameters involved in the problem, let us first revisit the results given in ~\citep{witten} where the deflection in the trajectory of a sub-relativistic spacecraft in the context of Planet Nine and the spacecraft two-body problem was studied. To understand how strong the gravitational effect of Planet Nine is on the spacecraft, kinetic energy per unit mass, $v^2/2$, and gravitational potential energy of Planet Nine, $-\mu_{P9}/\rho$, can be compared in the vicinity of Planet Nine. The current estimates on the mass and the orbit of Planet Nine given in ~\citep{BB3} suggest that the gravitational parameter is $\mu_{P9}\approx6.2\mu_{E}$ where $\mu_{E}=398600$ km$^3$$/$s$^2$ and semi-major axis is $a_{P9}\approx380$ AU. Then, the smallest impact parameter for the half sky search with 1000 spacecraft is $\rho\approx30$ AU from the method given in ~\citep{witten}. Let the speed of the spacecraft be $v=0.001c\approx63$ AU/year where $c=2.99792458\times10^5$ km/s is the speed of light. This sub-relativistic speed is chosen to make the mission duration reasonable. If the relevant parameters are put into the equations of the kinetic and potential energies, one can obtain that kinetic energy is eight orders of magnitude larger than the gravitational energy. So, the gravity of Planet Nine can be considered as a perturbative effect on a straight-line trajectory of a sub-relativistic spacecraft. In fact, the spacecraft follows a hyperbolic orbit with large eccentricity value and the trajectory can be considered as a distant planetary flyby. The reference frame and configuration of the two-body problem are given in Fig. \ref{fig:witten}. The straight black line is the unperturbed trajectory in the absence of Planet Nine, and the blue trajectory is the perturbed one. Finally, $\rho$ is the minimum distance to Planet Nine along the unperturbed trajectory. 

The position vector of the spacecraft in the vicinity of Planet Nine would take the form;
\begin{equation}
\vec{r}(t)=v_{0}t\hat{i}+\rho\hat{j}+\vec{r}_1(t),
\label{eq:deriv1}
\end{equation}    
where $\vec{r}_1(t)=x_1\hat{i}+y_1\hat{j}$ is the deviation from the straight line trajectory. For the unperturbed trajectory, the position at $t=0$ corresponds to the minimum distance between Planet Nine and the spacecraft. Then, the velocity and acceleration vectors become;
\begin{eqnarray}
\dot{\vec{r}}(t) & = & \left(v_{0}+\dot{x}_1\right)\hat{i}+\dot{y}_1\hat{j},   \label{eq:vel} \\
\ddot{\vec{r}}(t) & = & \ddot{x}_1\hat{i}+\ddot{y}_1\hat{j}.   \label{eq:accel} 
\label{eq:deriv2}
\end{eqnarray}   
The relative position vector of the spacecraft with respect to Planet Nine satisfies;
\begin{equation}
\ddot{\vec{r}}=-\frac{\mu_{P9}}{r^3}\vec{r},
\label{eq:deriv3}
\end{equation}    
where $r=\left[ (v_{0}t+x_1)^2+(\rho+y_1)^2 \right]^{1/2}$. Then, the components of the equations of motion are;
\begin{eqnarray}
\ddot{x}_1 & = & -\frac{\mu_{P9}}{\left( v_0^2t^2+2v_0tx_1+x_1^2+\rho^2+2\rho y_1+y_1^2\right)^{3/2}}\left(v_0 t+x_1\right),   \label{eq:accelx} \\
\ddot{y}_1 & = & -\frac{\mu_{P9}}{\left( v_0^2t^2+2v_0tx_1+x_1^2+\rho^2+2\rho y_1+y_1^2\right)^{3/2}}\left(\rho+y_1\right).   \label{eq:accely} 
\label{eq:deriv4}
\end{eqnarray}    
Since Planet Nine has a perturbative effect, the deviations would be small so that;
\begin{equation}
\frac{x_1}{v_0t}\ll1, \quad \frac{y_1}{\rho}\ll1.    
\label{eq:deriv5}
\end{equation}     
Then, the equations of motion take the following form;
\begin{equation}
\ddot{x}_1 =  -\frac{\mu_{P9}}{\left( v_0^2t^2+\rho^2 \right)^{3/2}}v_0 t,\quad 
\ddot{y}_1 =  -\frac{\mu_{P9}}{\left( v_0^2t^2+\rho^2 \right)^{3/2}}\rho.  
\label{eq:deriv6}
\end{equation}
which yield the perturbative velocities;
\begin{equation}
\dot{x}_1 =\frac{\mu_{P9}}{v_0\sqrt{v_0^2t^2+\rho^2}}+c_x,\quad
\dot{y}_1 =-\frac{\mu_{P9}}{\rho\sqrt{v_0^2t^2+\rho^2}}t+c_y,  
\label{eq:deriv7}
\end{equation}
where $c_x$ and $c_y$ are the integration constants to be found from the boundary conditions. As $t\rightarrow-\infty$, there should not be any gravitational effect of Planet Nine on the spacecraft, so the velocity should be;
\begin{equation}
\lim_{t\to-\infty} \dot{\vec{r}}(t)=v_0\hat{i}.
\label{eq:deriv8}
\end{equation} 
Thus, the proper boundary condition would be;
\begin{equation}
\lim_{t\to-\infty} \dot{\vec{r}}_1(t)=0.
\label{eq:deriv9}
\end{equation}
After applying the boundary conditions, the final forms of the perturbative velocities become;
\begin{equation}
\dot{x}_1 =\frac{\mu_{P9}}{v_0\sqrt{v_0^2t^2+\rho^2}},\quad
\dot{y}_1 =-\frac{\mu_{P9}}{\rho\sqrt{v_0^2t^2+\rho^2}}t-\frac{\mu_{P9}}{\rho v_0}.  
\label{eq:deriv10}
\end{equation}  
Finally, the magnitude of the transverse velocity change can be found as;
\begin{equation}
\Delta v_y = \bigg| \lim_{t\to\infty} \dot{y}_1 - \lim_{t\to-\infty} \dot{y}_1 \bigg|= \bigg| \lim_{t\to\infty} \dot{y}_1 \bigg| = \frac{2\mu_{P9}}{\rho v_0}.
\label{eq:deriv11}
\end{equation}
Here, we focus on the measurement of transverse displacement, instead of time delay in the signal to relax the accurate timekeeping requirement. So, the maximum angular displacement is given as~\citep{bruteforce};
\begin{equation}
\alpha = \frac{\Delta v_y}{v_x^{(0)}}.
\label{eq:deriv12}
\end{equation}
Using (\ref{eq:deriv11}) in (\ref{eq:deriv12}), one has 
\begin{equation}
\alpha = \frac{2\mu_{P9}}{\rho v_0^2} \approx 1\times10^{-8}\,\text{radians}.
\label{eq:deriv13}
\end{equation}
In~\citep{bruteforce}, it was noted that that the angular displacement could be measured precisely with the use of VLBI with a detection threshold of order $10^{-9}$ radians for high-frequency sources. (\ref{eq:deriv13}) is already above the threshold. To further improve the detectability, the impact parameter may be decreased by increasing the number of spacecraft or spacecraft may be slowed down.  
\section{The Sun-Planet Nine-Spacecraft system: a three-body problem} 
In this section, the angular displacement is studied in the context of the Sun--Planet Nine-spacecraft three-body problem to understand whether the three-body context provides new prospects for the detection. To have an idea of the effect of the Sun, the ratio of gravitational acceleration due to Planet Nine and the Sun is calculated by taking the gravitational parameter of the Sun as $\mu_{S}=132712\times10^6$ km$^3$$/$s$^2$;
\begin{equation}
\left(\frac{\mu_{S}}{a_{P9}^2}\right)/\left(\frac{\mu_{P9}}{\rho^2}\right)\approx335,
\label{eq:sun}
\end{equation}
moreover, the tidal acceleration due to the Sun and Planet Nine on the spacecraft has the ratio
\begin{equation}
\left(\frac{\mu_{S}}{a_{P9}^3}\right)/\left(\frac{\mu_{P9}}{\rho^3}\right)\approx26.
\label{eq:sun2}
\end{equation}
So these motivate us to consider the effect of the Sun on the trajectory of the spacecraft.

For the three-body system, we shall adopt the Circular Restricted Three-Body Problem (CRTBP) as the underlying dynamical model. In using this model, the following assumptions are made:
\begin{itemize}
\item The Sun and Planet Nine are moving in a circular orbit around their barycenter. In fact, the orbit of Planet Nine was estimated to be elliptic ($e\approx0.2-0.5$)~\citep{Batygin2019}, but the simple circular model may still provide insight into the effect of the Sun.
\item The orbit of Planet Nine was estimated as moderately inclined ($i\approx15^{\circ}-25^{\circ}$)~\citep{Batygin2019}. However, we use the approximation that the Sun, Planet Nine, and spacecraft are moving in the same plane to reduce the dimension of the problem.
\item The mass of the spacecraft is negligible such that it does not affect the motion of Sun and Planet Nine, which is the case in our problem.
\end{itemize}
The equations of motion for the planar CRTBP of Sun--Planet Nine-spacecraft are given as follows~\citep{curtis};
\begin{eqnarray}
\ddot{x} - 2\Omega\dot{y} - \Omega^{2}x & = & -\frac{\mu_{S}}{r_{1}^{3}}  \left(x + \pi_{2}r_{12} \right) - \frac{\mu_{P9}}{r_{2}^{3}}  \left(x - \pi_{1}r_{12}\right),  \\
\ddot{y} + 2\Omega\dot{x} - \Omega^{2}y & = & -\frac{\mu_{S}}{r_{1}^{3}}y - \frac{\mu_{P9}}{r_{2}^{3}}y, 
\label{eq:CRTBP}
\end{eqnarray}
where $\pi_{1}=m_S/(m_S+m_{P9})$ and $\pi_{2}=m_{P9}/(m_S+m_{P9})$ are the mass ratios, $r_{12}=a_{P9}$ is the average distance between the Sun and Planet Nine, $\Omega=\sqrt{\frac{(\mu_{S}+\mu_{P9})}{r_{12}^{3}}}$ is the angular velocity of the Sun--Planet Nine rotating frame, lastly $r_1$ and $r_2$ are the distances of the Sun-spacecraft and Planet Nine-spacecraft, respectively. 

We solve the equations of motion of CRTBP with numerical integration for the particular initial conditions. In numerical integration, we use the MATLAB built-in ode45 solver with absolute and relative tolerances of $10^{-13}$. It is important to note that the transverse displacement should be calculated relative to the inertial frame, although the motion is obtained in the rotating frame of CRTBP. Therefore, frame transformations between the rotating and the inertial frames are applied to obtain the angular displacement. Regarding the duration of integration, we assume that measurement starts at a distance to the Planet Nine where the analytical and numerical results are compatible. This distance is taken as 150 AU and the duration of measurement is calculated twice the time required to reach Planet Nine. For instance, if the spacecraft reaches Planet Nine in 1 years from start of the measurement, then the total integration duration would be 2 years. The symmetric time interval is assumed in the calculation of transverse displacement.

\begin{figure}
\centering
\includegraphics[width=.9\textwidth]{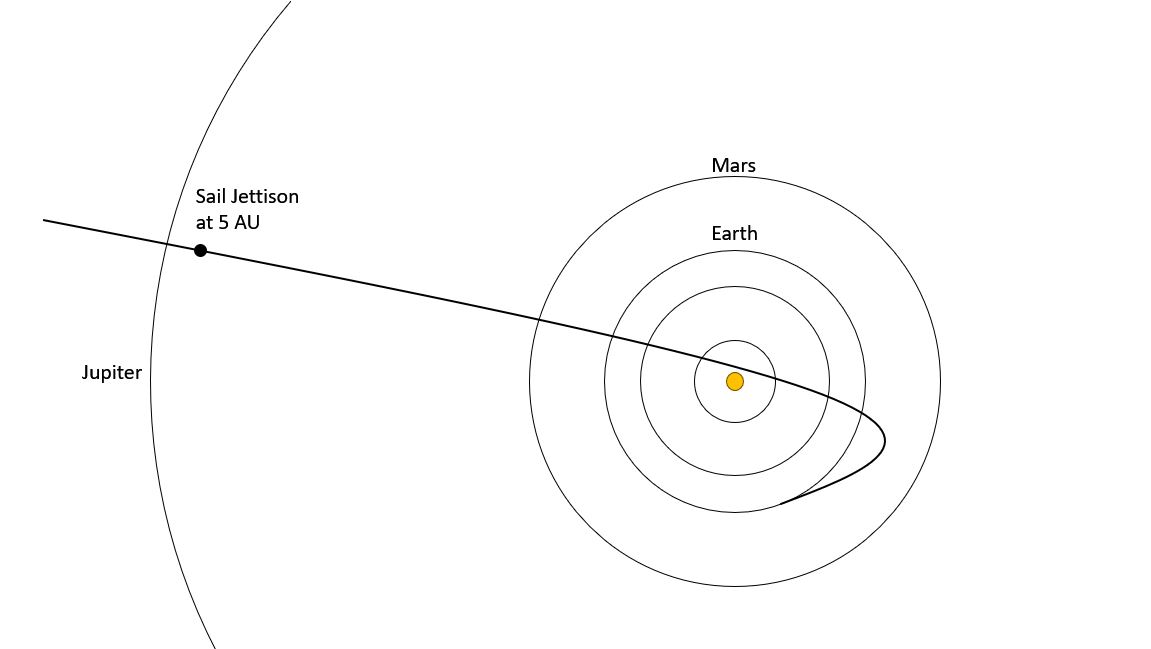}
\caption{An example of an interstellar escape trajectory with solar sail (redrawn along the lines of the Figure 13 in~\citep{sauer}).}
\label{fig:interstellar}
\end{figure}

\begin{figure}
\centering
\includegraphics[width=.9\textwidth]{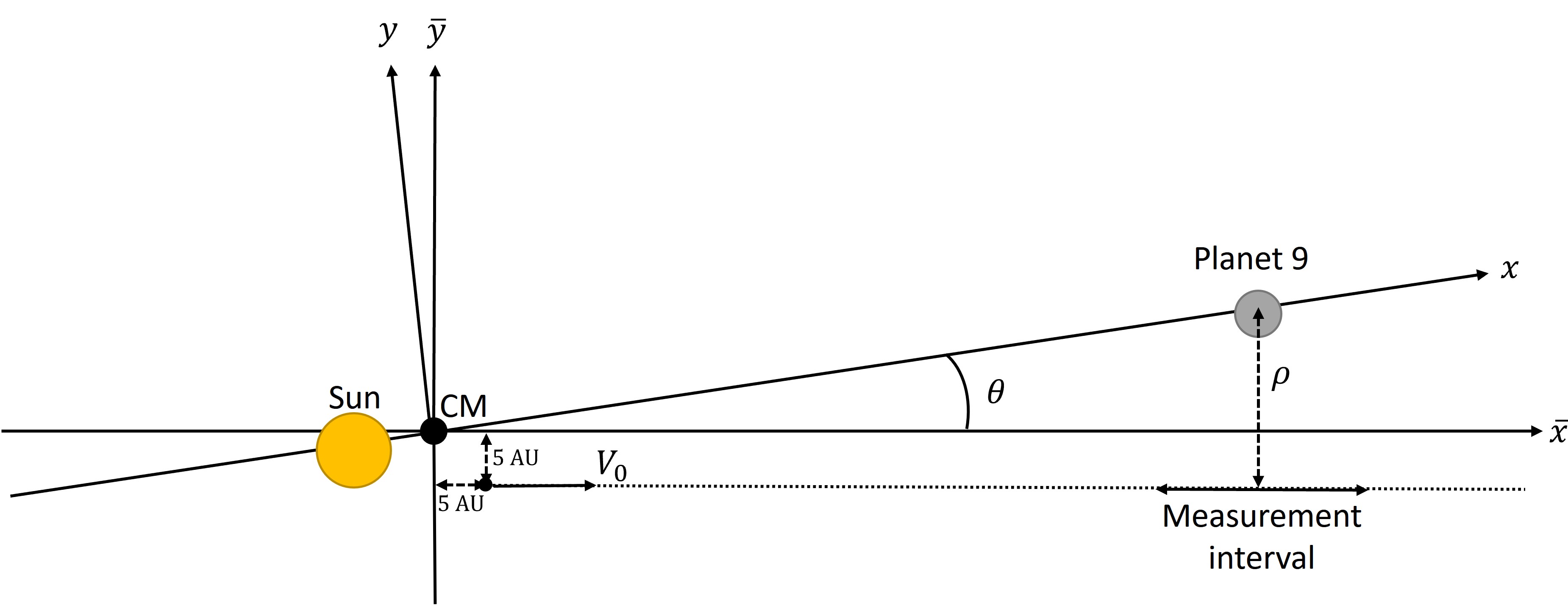}
\caption{$\bar{x}$-$\bar{y}$ is an arbitrary inertial frame and $x-y$ is the rotating frame of Sun--Planet Nine-spacecraft CRTBP. The spacecraft is sent horizontally with a speed $V_0$, and the initial position of Planet Nine makes an angle $\theta$ i.e. calculated from the desired impact parameter $\rho$. More explicitly, the initial position vector is $\vec{r}_0 = 5\hat{\bar{x}}-5\hat{\bar{y}} \ [AU]$ , the initial velocity vector is $\vec{v}_0 = V_0\hat{\bar{x}}$, and $\theta=\arcsin{(r_y/r_{P9})}$. This initial configuration is named as ``case A''.}
\label{fig:frame1}
\end{figure}

\begin{figure}
\centering
\includegraphics[width=.9\textwidth]{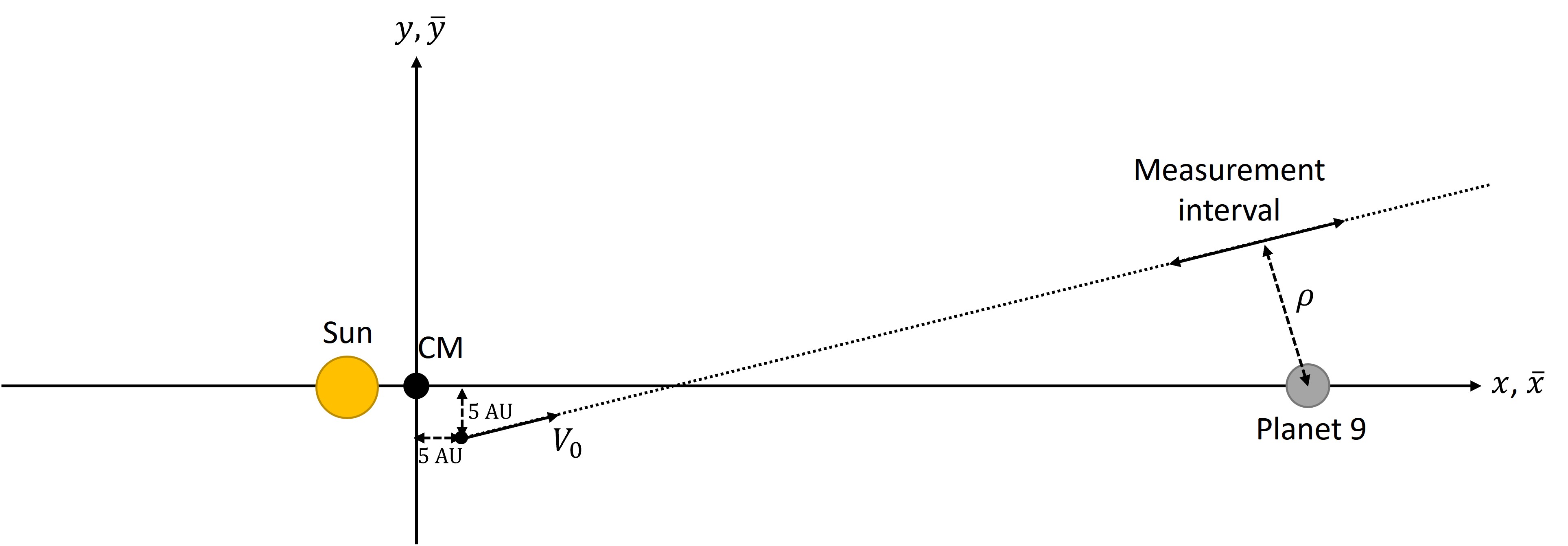}
\caption{$\bar{x}$-$\bar{y}$ is an arbitrary inertial frame and $x-y$ is the rotating frame of Sun--Planet Nine-spacecraft CRTBP. Spacecraft is sent diagonally with a speed $V_0$, and Planet Nine is initially on the $x$-axis. More explicitly, the initial position vector is $\vec{r}_0 = 5\hat{\bar{x}}-5\hat{\bar{y}} \ [AU]$ and the direction of the velocity vector of the spacecraft is calculated from the basic trigonometric relations to achieve desired impact parameter $\rho$. This initial configuration is named as ``case B''.}
\label{fig:frame2}
\end{figure}

In order to decide on the proper initial conditions, escape trajectories from the Solar System are examined. In~\citep{vulpetti,sauer}, interstellar mission trajectories with the help of solar sails were studied. Due to the reduction of the solar flux and its propulsive effect, in~\citep{sauer} jettison of the solar sail is estimated to be at approximately 5 AU. An example of an interstellar escape trajectory with a solar sail is given in Fig. \ref{fig:interstellar}. In our analysis, we assume that the spacecraft reaches the sub-relativistic speed at 5 AU, so the initial position for the numerical integration is chosen by considering this assumption. The two possible initial configurations are considered which are given in Fig. \ref{fig:frame1} and Fig. \ref{fig:frame2}.

\begin{figure}
\centering
\includegraphics[width=.9\textwidth]{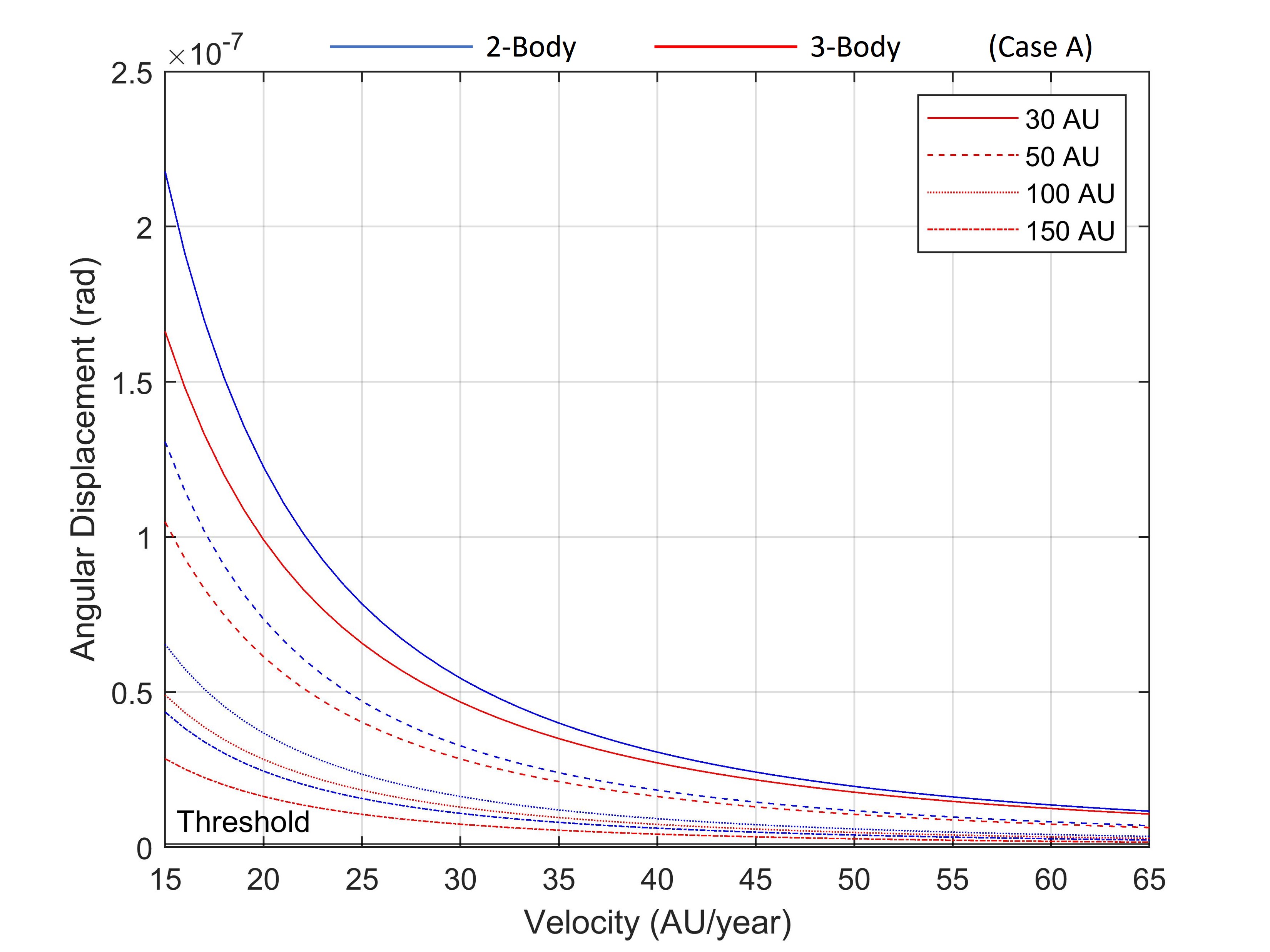}
\caption{Angular displacement of two-body and three-body trajectories with different velocities and impact factors for the Case A. The two-body results are obtained from the Planet Nine and spacecraft two-body problem, and the three-body results are obtained from the difference between Sun--Planet Nine-spacecraft three-body trajectory and Sun-spacecraft two-body trajectory.}
\label{fig:3bodyCaseA}
\end{figure}

\begin{figure}
\centering
\includegraphics[width=.9\textwidth]{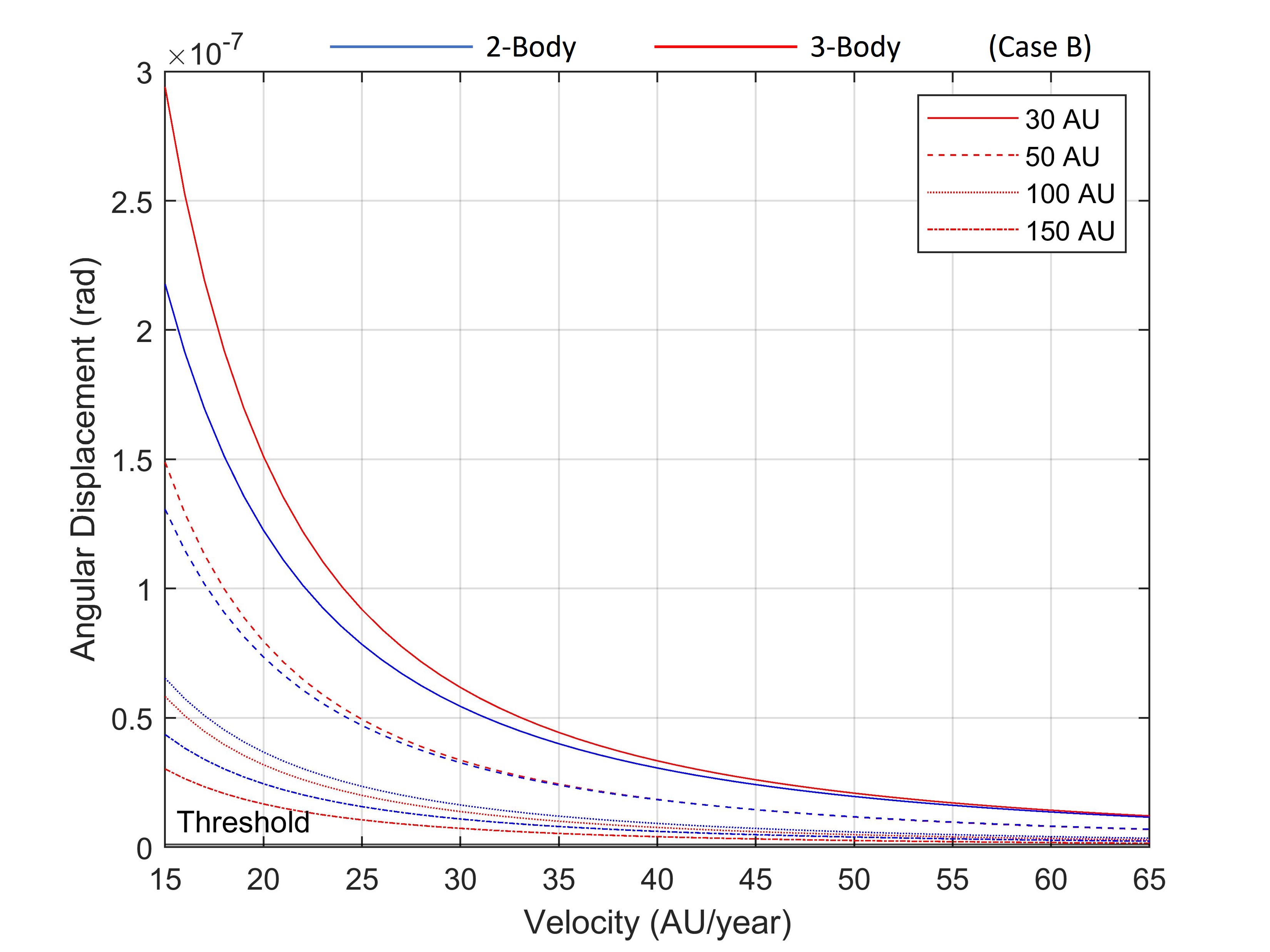}
\caption{Angular displacement of two-body and three-body trajectories with different velocities and impact factors for the Case B. The two-body results are obtained from the Planet Nine and spacecraft two-body problem, and the three-body results are obtained from the difference between Sun--Planet Nine-spacecraft three-body trajectory and Sun-spacecraft two-body trajectory.}
\label{fig:3bodyCaseB}
\end{figure}

In the presence of the Sun, the unperturbed or nominal trajectory of the spacecraft is defined as the trajectory obtained from the Sun-spacecraft two-body problem given as;
\begin{equation}
\ddot{\vec{r}}=-\frac{\mu_{S}}{r^{3}}\vec{r},
\label{eq:2body}
\end{equation}
where $\vec{r}$ is the position vector of the spacecraft relative to the Sun. Then, perturbation due to Planet Nine is calculated as the difference between the Sun--Planet Nine-spacecraft three-body trajectory obtained from (\ref{eq:CRTBP}); and the Sun-spacecraft two-body trajectory obtained from Eqn. (\ref{eq:2body}). The results of both initial configurations are given for different velocities and impact parameters in Fig. \ref{fig:3bodyCaseA} and Fig. \ref{fig:3bodyCaseB}.

In the figures, two-body results correspond to the perturbation of Planet Nine on a straight-line trajectory and are obtained from (\ref{eq:deriv13}). The three-body results correspond to the perturbation of Planet Nine on a trajectory that is defined by the Sun and spacecraft two-body problem. The two-body and three-body results are compared to assess whether considering the Sun yields a significant effect on the detection of Planet Nine. Besides, different velocities and impact parameters are also considered since the angular displacement depends on those parameters. Velocities are decreased until $15$ AU/year because further reduction results in a longer mission duration; and different impact parameters are used for the case of different number of spacecraft.

In Fig. \ref{fig:3bodyCaseA}, results for the initial configuration of case A are given. The angular displacement due to Planet Nine is larger in the two-body context for all the impact parameters and velocities. In CRTBP, Planet Nine and the Sun are moving in circular orbits, unlike the two-body setting in which one considers Planet Nine as stationary. Due to the movement of Planet Nine, the distance between Planet Nine and the spacecraft is larger at the flyby for the three-body trajectory. So, a smaller angular displacement is obtained. As the velocity decreases, the kinetic energy becomes comparable with the potential energy due to the Sun and Planet Nine. Therefore, their effects become visible, and larger displacement occurs. As the impact parameter increases, spacecraft flybys with Planet Nine at a larger distance. So, its gravitational effect decreases and a smaller displacement occurs. Lastly, all the angular displacements are above the detection threshold.

In Fig. \ref{fig:3bodyCaseB}, results for the initial configuration of case B are given. The angular displacement due to Planet Nine is larger in the three-body context for the impact parameters of 30 AU and 50 AU. Due to the movement of Planet Nine in the CRTBP, the distance between Planet Nine and the spacecraft is smaller at the flyby. So, the angular displacement due to Planet Nine is larger. But two-body results are still larger for the impact parameters of 100 AU and 150 AU. To increase the angular displacement in the three-body context, one can increase the measurement duration. All the angular displacements are above the detection threshold.

To sum up, we found that the effect of the Sun is not directly seen in the detection but the presence of the Sun causes the motion of Planet Nine and this yields larger or smaller angular displacements depending on the initial condition and measurement duration.    
\section{Planetary and Kuiper Belt Effects on the Spacecraft Trajectory}
\begin{figure}
\centering
\includegraphics[width=.9\textwidth]{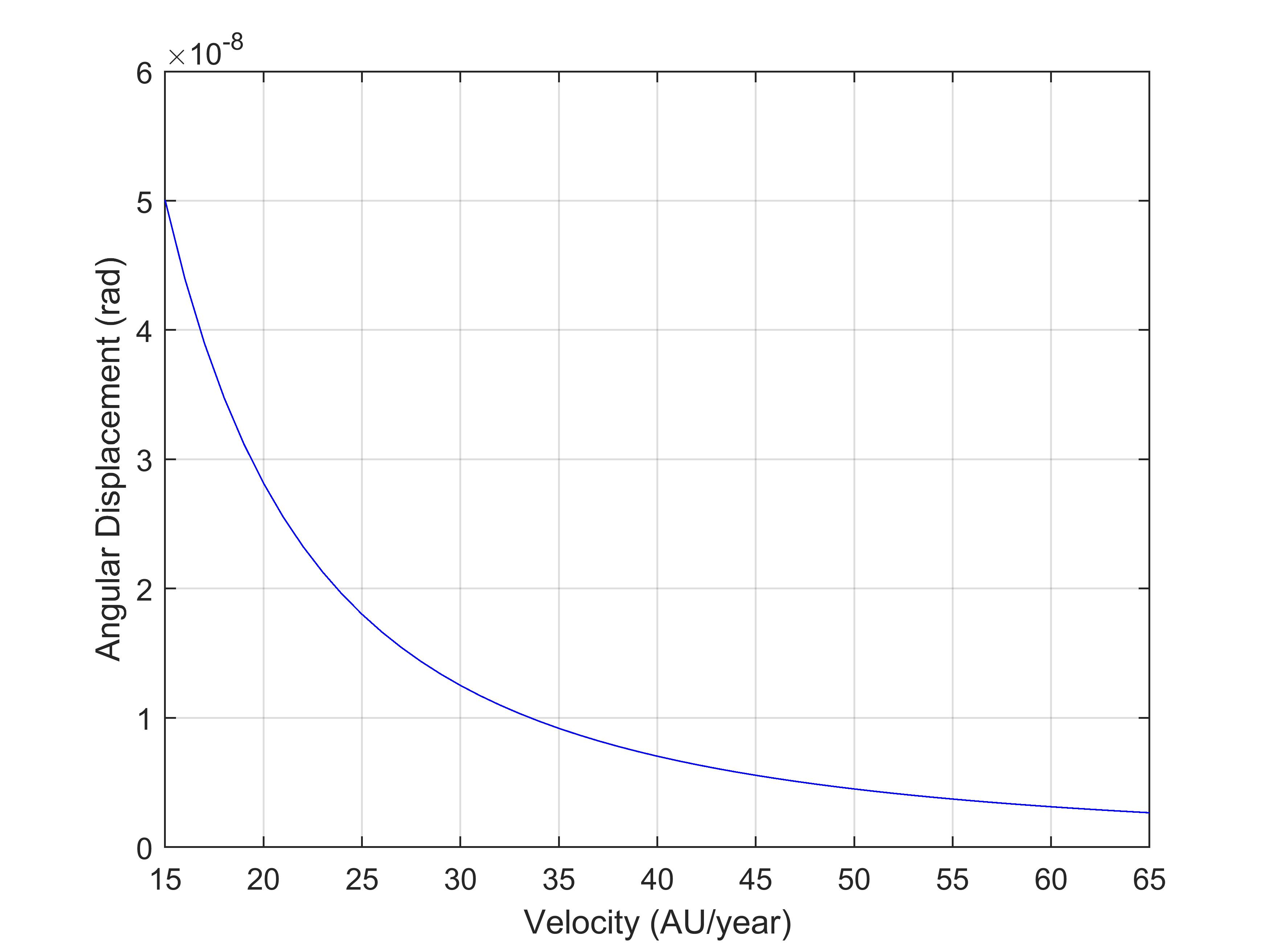}
\caption{Angular displacement due to outer planets, namely Jupiter, Saturn, Uranus and Neptune, with different velocities for the initial configuration of case B.}
\label{fig:planets}
\end{figure}

We first consider the gravitational effects of outer planets, namely Jupiter, Saturn, Uranus and Neptune, on the spacecraft trajectory. Since the start of the measurement is sufficiently far away from the planets, the motion of planets does not effect the displacement at the leading order. Therefore, we consider the planets as stationary masses placed at the relevant orbital radius. The initial configuration of case B is considered since it has a larger angular displacement compared to case A. The perturbative effect of planets is investigated as the difference between the Sun--Planets-spacecraft trajectory and the Sun-spacecraft two-body trajectory. In the analysis, the effects of outer planets are considered simultaneously. The physical and orbital properties of the planets are taken from~\citep{curtis}. The resulting angular displacement versus velocity plot is given in Fig. \ref{fig:planets}.  
 
As the velocity increases, the effects of outer planets decrease. For the lower velocities, spacecraft is exposed to the gravity of planets in a longer time interval and therefore larger displacement occurs. However, the effect of outer planets is about one order smaller than the effect of Planet Nine.

Next, we investigate the effect of Kuiper Belt on the spacecraft trajectory and modeled the Kuiper Belt as a ring with uniform mass distribution. The force due to the gravitational field of a ring on the plane that ring is placed is given as~\citep{michal};
\begin{equation}
f(r)=-\frac{\kappa m M}{\pi R^2} \frac{\left(1+\frac{r}{R}\right)E\left(\frac{4 \left(\frac{r}{R}\right)}{\left(1+\frac{r}{R}\right)^2}\right) +\left(1-\frac{r}{R}\right)K\left(\frac{4 \left(\frac{r}{R}\right)}{\left(1+\frac{r}{R}\right)^2}\right)}{\frac{r}{R}\left[\left(\frac{r}{R}\right)^2-1\right]}
\label{eq:kuip}
\end{equation}
where $K$ and $E$ are the complete elliptic integrals of the first and second kind, $\kappa$ is the gravitational constant, $m$ and $M$ are the masses of test particle and ring, respectively, $r$ is the distance of test particle to the center of the ring and $R$ is the radius of the ring.

In calculating the force due to Kuiper Belt, the average radius ($R$) and the mass ($M$) of the Kuiper Belt is taken from~\citep{pitjeva} as $R=43.6$ AU and $M=1.97\times10^{-2}M_{Earth}$. Then, the ratio of gravitational force due to Planet Nine and Kuiper Belt for $r=380$ AU and $\rho=30$ AU is;
\begin{equation}
\frac{F_{P9}}{F_{Kuiper}}\approx10^{5}
\label{eq:forceCom}
\end{equation}
So, the effect of Kuiper Belt is negligible.
 
\section{Drag Force on the Spacecraft Trajectory} 
Due to the interstellar gas particles and dust, the spacecraft is exerted by a drag force~\citep{loeb1};
\begin{equation}
F_{drag}=1.4n_{H}m_{H}v^2A_{sp},
\label{eq:drag}
\end{equation}
where $n_H$ is the proton number density and taken as $n_H=1$ cm$^{-3}$ which is a standard parameter for the interstellar medium~\citep{loeb1}, $m_H$ is the proton mass, $v$ is the speed and $A_{sp}$ is the drag area. Then, the equation of motion of the spacecraft with drag force is given as;
\begin{equation}
\ddot{\vec{r}}=-\frac{\mu_{S}}{r^{3}}\vec{r}-\frac{F_{drag}}{M_{sp}}\frac{\vec{v}}{|\vec{v}|}.
\label{eq:drageom}
\end{equation}
where $M_{sp}$ is the spacecraft mass. The first term represents the nominal trajectory of the spacecraft that is defined by the Sun and the spacecraft two-body problem. In calculating the drag force, the direction is opposite to that of the velocity. We assume that the drag force affects the trajectory of the spacecraft all the time during the cruise with the same proton number density. 

\begin{figure}
\centering
\includegraphics[width=.9\textwidth]{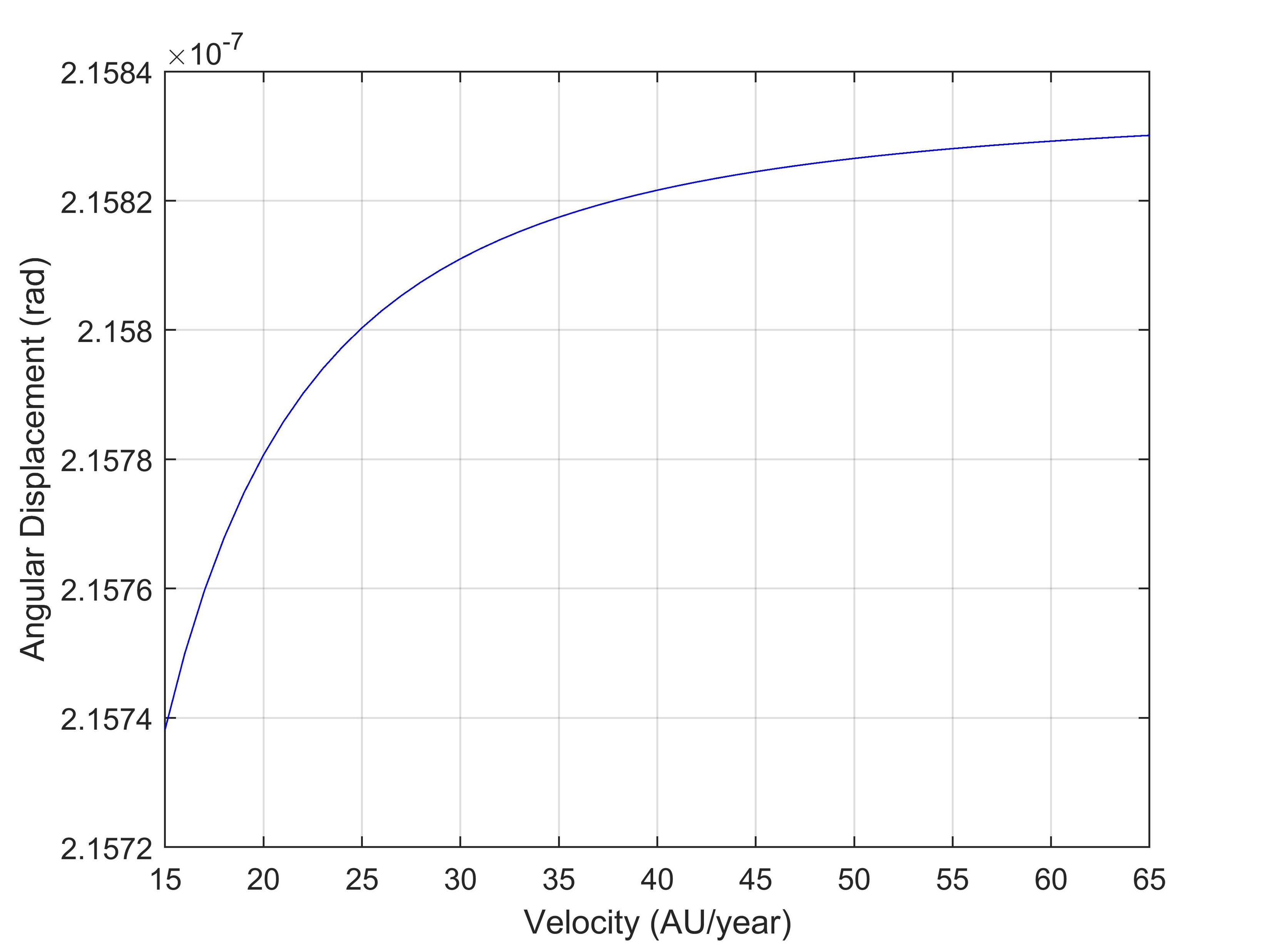}
\caption{Angular displacement due to drag force with different velocities for a spacecraft with drag area $A_{sp}=120$ m$^2$ and mass $M_{sp}=5.45$ kg and for the initial configuration of case B.}
\label{fig:drag}
\end{figure}

In analyzing the effect of drag force, the initial configuration of case B is considered since it has a larger angular displacement compared to case A. The perturbative effect of the drag force is investigated as the difference between the Sun-Drag-spacecraft trajectory and the Sun-spacecraft two-body trajectory. In calculating the drag force, we take the spacecraft parameters according to a possible technology demonstration mission with solar sail given in~\citep{solarsail2} as $A_{sp}=120$ m$^2$ and $M_{sp}=5.45$ kg. The resulting angular displacement versus velocity plot is given in Fig. \ref{fig:drag}.

As the velocity increases, drag also increases and larger angular displacement is obtained. The order of displacement due to gravity of Planet Nine for $\rho=30$ AU is slightly larger than the drag force for the lower velocities but still to distinguish effect of Planet Nine from drag force would be hard.  

\section{Magnetic Force on the Spacecraft Trajectory} 
\begin{figure}
\centering
\includegraphics[width=.9\textwidth]{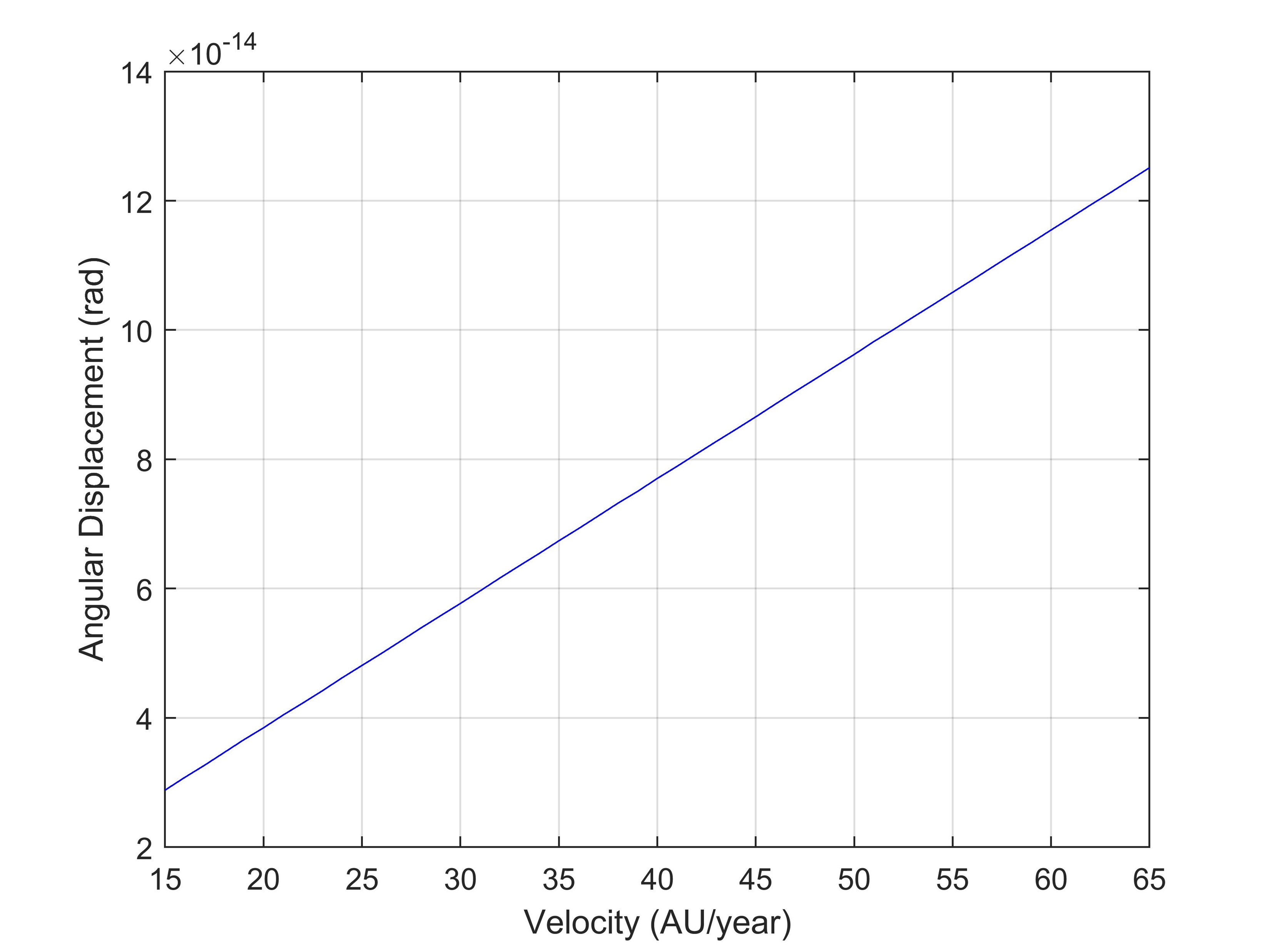}
\caption{Angular displacement due to magnetic force with different velocities for a spacecraft with mass $M_{sp}=5.45$ kg and $\varrho=1.42$ gram cm$^{-3}$ and for the initial configuration of case B}
\label{fig:magnetic}
\end{figure}

During the cruise, the spacecraft gets electrically charged due to colliding interstellar particles and also due to photoelectric effects from solar and interstellar photons~\citep{loeb1}. Then, the interstellar magnetic field exerts a magnetic force on the charged spacecraft and curves the trajectory. The magnetic force due to interstellar medium is given as~\citep{loeb1};
\begin{equation}
F_{mag}=\frac{e}{c}Z_{sp}vB_{\perp},
\label{eq:magnetic1}
\end{equation}
in Gaussian units. Here, $e$ is the elementary charge, $v$ is the speed, $B_{\perp}$ is the magnetic field component perpendicular to the motion, $c$ is the speed of light and $Z_{sp}$ is the maximum charge of the spacecraft that is given as~\citep{loeb1};
\begin{equation}
Z_{sp}=\frac{m_ev^2}{e^2}\left(\frac{M_{sp}}{\varrho}\right)^{1/3},
\label{eq:magnetic2}
\end{equation}
where $m_e$ is the mass of the electron, $M_{sp}$ is the mass of the spacecraft and $\varrho$ is the density of spacecraft. One can look at~\citep{loeb2} for a detailed discussion about the calculation of $Z_{sp}$. Then, the equation of motion of the spacecraft with the magnetic force is given as;
\begin{equation}
\ddot{\vec{r}}=-\frac{\mu_{S}}{r^{3}}\vec{r}+\frac{F_{mag}}{M_{sp}}\hat{b},
\label{eq:mageom}
\end{equation}
where the direction $\hat{b}$ is perpendicular to both the velocity vector and the orbit angular momentum vector of the spacecraft so that the maximum displacement occurs due to the magnetic force. The magnitude of the magnetic field is taken from~\citep{loeb1} as $B_{\perp}\sim5 \mu G$, and the spacecraft parameters are determined according to a possible technology demonstration mission with solar sail given in~\citep{solarsail2} as $M_{sp}=5.45$ kg and $\varrho=1.42$ gram cm$^{-3}$. Here again, the initial configuration of case B is considered. Then, the perturbative effect of the magnetic force is investigated as the difference between the Sun-Magnetic Force-spacecraft trajectory and the Sun-spacecraft two-body trajectory. The resulting angular displacement versus velocity plot is given in Fig. \ref{fig:magnetic}.

As the velocity increases, the magnetic force also increases and curves the trajectory. Therefore, a larger angular displacement is obtained for the high velocities. The resulting angular displacement due to magnetic force is below the detection threshold and is not detectable for the parameters used in the analysis. 

\section{Solar Radiation Pressure on the Spacecraft Trajectory} 
\begin{figure}
\centering
\includegraphics[width=.9\textwidth]{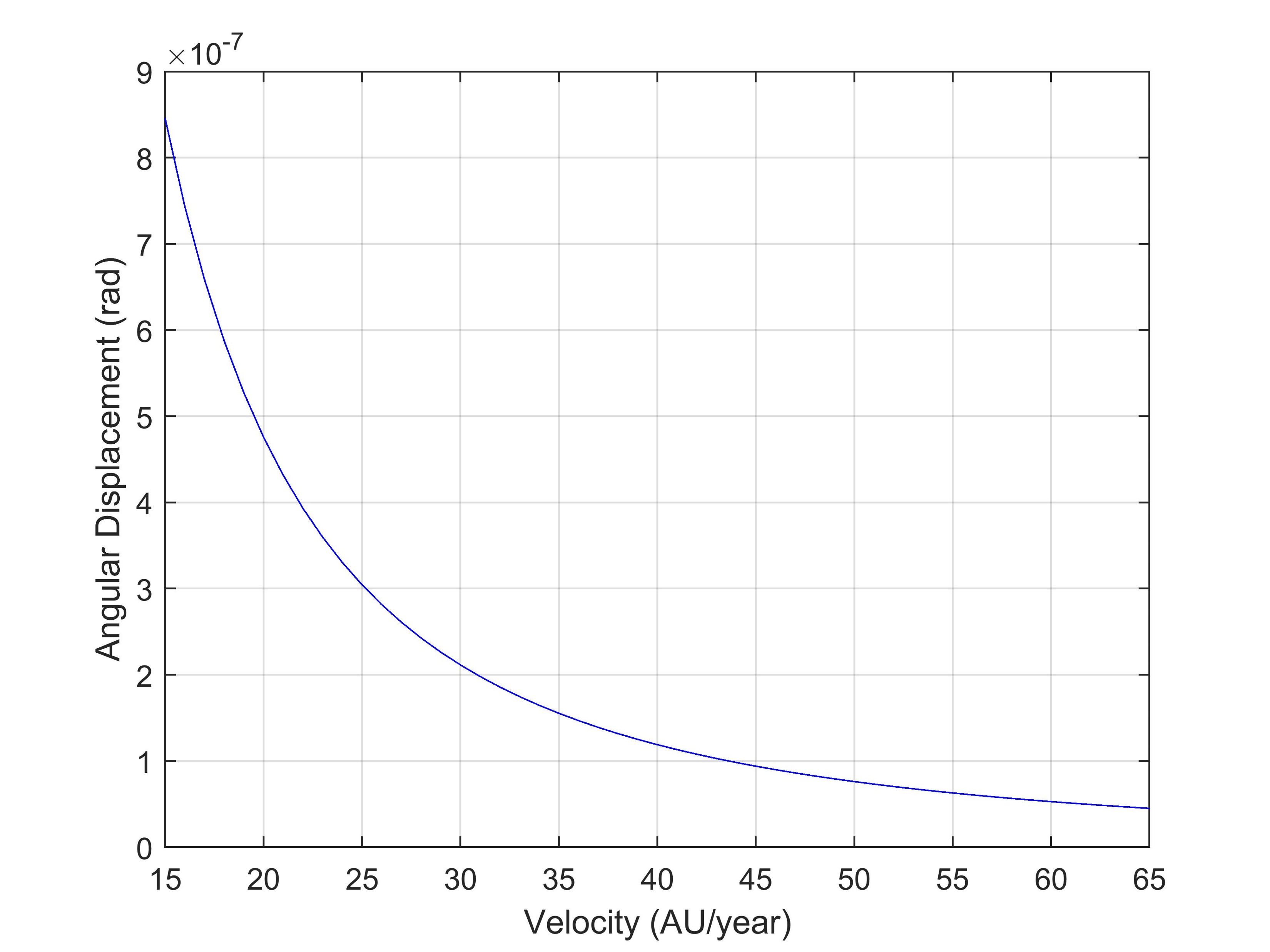}
\caption{Angular displacement due to SRP with different velocities for a spacecraft with mass $M_{sp}=5.45$ kg and SRP area $A_{S}=120$ m$^2$ and for the initial configuration of case B.}
\label{fig:SRP}
\end{figure}

The spacecraft is exposed to the Solar Radiation Pressure (SRP) during the cruise which perturbs the trajectory. The acceleration due to SRP on the spacecraft is given as~\citep{curtis};
\begin{equation}
F_{srp}=\nu\frac{S}{c}C_{R}A_{S},
\label{eq:srp}
\end{equation}
where $\nu$ is the eclipse factor that is always 1 in our analysis due to the assumption that the spacecraft is always in sunlit, $c$ is the speed of light, $A_{S}$ is the surface area of the spacecraft that is exposed to SRP, $C_{R}$ is the reflectivity coefficient of the spacecraft surface and takes a moderate value $C_{R}=1.5$. Lastly, $S$ is the solar flux as a function of the distance of the spacecraft that is given as~\citep{curtis};
\begin{equation}
S=S_0\left(\frac{R_0}{R}\right)^2,
\label{eq:S}
\end{equation}
where $S_0=63.15\times10^6$ W/m$^2$ is the radiated power intensity of the photosphere, $R_0=696000$ km is the radius of the photosphere and $R$ is the distance between the Sun and the spacecraft. Then, the equation of motion of the spacecraft with SRP is given as;
\begin{equation}
\ddot{\vec{r}}=-\frac{\mu_{S}}{r^{3}}\vec{r}+\frac{F_{srp}}{M_{sp}}\hat{r},
\label{eq:srpeom}
\end{equation}
where $\hat{r}$ is the unit vector from Sun to spacecraft. In calculating the SRP, the SRP area $A_{S}=120$ m$^2$ is taken as the same as the drag area, and the mass of the spacecraft is taken as $M_{sp}=5.45$ kg. Here again, the initial configuration of case B is considered. The perturbative effect of SRP is investigated as the difference between the Sun-SRP-spacecraft trajectory and the Sun-spacecraft two-body trajectory. The resulting angular displacement versus velocity plot is given in Fig. \ref{fig:SRP}.

As the velocity increases, the spacecraft is exposed to the SRP in a shorter time interval and therefore the angular displacement decreases. We found that the order of angular displacement due to SRP is larger than the gravity of Planet Nine.
\section{General Relativistic Effects on the Spacecraft Trajectory}

\subsection{Frame-Dragging Effect}
Due to the rotation of the central massive body, a frame-dragging effect arises in General Relativity. Here, we first study the effect of frame-dragging on the detection of Planet Nine by comparing its magnitude with that of the gravity of Planet Nine. Frame-dragging correction to the acceleration of the spacecraft is given as~\citep{sosnica};
\begin{equation}
\Delta{\ddot{\vec{r}}_{FD}} = (1+\gamma)\frac{GM}{c^2r^3}\left[\frac{3}{r^2}\left(\vec{r}\times\dot{\vec{r}}\right)\left(\vec{r}\cdot\vec{J}\right)+\left(\dot{\vec{r}}\times\vec{J}\right)\right],
\label{eq:framedragging}
\end{equation}
where $\gamma$ is a post-Newtonian parameter which we will take to be 1, $GM$ is the gravitational parameter, $c$ is the speed of light, $\vec{r}$ and $\dot{\vec{r}}$ are the position and velocity vectors of the spacecraft and lastly $\vec{J}$ is the angular momentum vector per unit mass of the relevant body. 

We want to investigate the frame-dragging effect of our Solar System on a spacecraft with $v=0.001c$ at the distance of Planet Nine. So the parameters are taken as follows;
$GM$ is the gravitational parameter of the Sun, $r$ is the average distance between the Sun and Planet Nine which is $a_{P9}$, and lastly $\vec{J}$ is the orbit angular momentum vector per unit mass of Jupiter since it has the major contribution to the angular momentum of our Solar System. Also, since we study the planar motion, position and velocity vectors of the spacecraft, i.e. $\vec{r}$ and $\dot{\vec{r}}$, would be perpendicular to the angular momentum vector $\vec{J}$. In the vicinity of Planet Nine, i.e. at a distance of $\rho$, the ratio of the Newtonian gravity of Planet Nine ($|\ddot{\vec{r}}_{P9}|$) and the frame-dragging effect due to the Solar System ($|\Delta{\ddot{\vec{r}}_{FD,SS}}|$) is
\begin{equation}
\frac{|\ddot{\vec{r}}_{P9}|}{|\Delta{\ddot{\vec{r}}_{FD,SS}}|} \approx 10^{6}.
\label{eq:ratio1}
\end{equation}
So, the frame-dragging effect due to the Solar System is negligible. 

Next, we look at the frame-dragging effect due to Planet Nine on the spacecraft. The specific orbit angular momentum of Planet Nine can be calculated by considering the orbital elements given in~\citep{Batygin2019} as follows~\citep{curtis};
\begin{equation}
J_{P9} = \sqrt{a_{P9}\mu_{P9}(1-e_{P9}^2)},
\label{eq:hmag}
\end{equation}
where $e_{P9}$ is the eccentricity and is chosen to be $e_{P9}=0.2$ to have a larger angular momentum. In the vicinity of Planet Nine, the frame-dragging effect due to the orbit angular momentum of Planet Nine on the spacecraft ($|\Delta{\ddot{\vec{r}}_{FD,P9}}|$) can be calculated by considering $GM=\mu_{P9}$, $r=\rho=30$ AU, $v=0.001c$ and $J=J_{P9}$. Its comparison with the Newtonian gravity of Planet Nine is given as follows;  
\begin{equation}
\frac{|\ddot{\vec{r}}_{P9}|}{|\Delta{\ddot{\vec{r}}_{FD,P9}}|} \approx 10^{9}.
\label{eq:ratio2}
\end{equation}
So, the frame-dragging effect due to orbit angular momentum of Planet Nine is negligible.

Lastly, we consider Planet Nine as an extremal black hole, so the maximum spin angular momentum per unit mass of a black hole can be given as;
\begin{equation}
J_{BH} = \frac{GM}{c}, 
\label{eq:bh}
\end{equation}
where $GM=\mu_{P9}$. In the vicinity of Planet Nine, the frame-dragging effect due to the spin angular momentum of a black hole on the spacecraft ($|\Delta{\ddot{\vec{r}}_{FD,BH}}|$) can be calculated by considering $GM=\mu_{P9}$, $r=\rho=30$ AU, $v=0.001c$ and $J=J_{BH}$. Its comparison with the Newtonian gravity of Planet Nine is given as follows; 
\begin{equation}
\frac{|\ddot{\vec{r}}_{P9}|}{|\Delta{\ddot{\vec{r}}_{FD,BH}}|} \approx 10^{16}.
\label{eq:ratio3}
\end{equation}
So, the frame-dragging effect due to spin angular momentum of a black hole that equals the mass of Planet Nine is negligible. It has the smallest magnitude among the other frame-dragging effects.

\subsection{Schwarzschild Effect}
General Relativistic corrections to the Newtonian gravity, the so-called Schwarzschild effect, on the spacecraft can be given as~\citep{sosnica};
\begin{equation}
\Delta{\ddot{\vec{r}}_{Sch}} = \frac{GM}{c^2r^3}\Biggl\{
\left[2(\beta+\gamma)\frac{GM}{r}-\gamma\dot{\vec{r}}\cdot\dot{\vec{r}}\right]\vec{r}+
2(1+\gamma)(\vec{r}\cdot\dot{\vec{r}})\dot{\vec{r}}\Biggl\},
\label{eq:sch}
\end{equation}
where $\beta$ and $\gamma$ are the post-Newtonian parameters which we set to 1, $GM$ is the gravitational parameter, $c$ is the speed of light, $\vec{r}$ and $\dot{\vec{r}}$ are the position and velocity vectors of the spacecraft.

The Schwarzschild effect due to the Sun on the spacecraft ($|\Delta{\ddot{\vec{r}}_{Sch,S}}|$) in the vicinity of Planet Nine is calculated by considering $GM=\mu_{S}$, $r=a_{P9}$, $v=0.001c$ and by assuming $\vec{r}$ and $\dot{\vec{r}}$ are in the same direction so that $\vec{r}\cdot\dot{\vec{r}}=|\vec{r}||\dot{\vec{r}}|$. Its comparison with the Newtonian gravity of Planet Nine is given as follows; 
\begin{equation}
\frac{|\ddot{\vec{r}}_{P9}|}{|\Delta{\ddot{\vec{r}}_{Sch,S}}|} \approx 10^{3}.
\label{eq:ratio4}
\end{equation}
The Schwarzschild effect due to Sun is larger than the frame-dragging effects. But still, the deflection on the spacecraft trajectory due to the Newtonian gravity of Planet Nine can be distinguishable from the Schwarzschild effect of the Sun since there are three order differences. 

Next, we look for the Schwarzschild effect due to Planet Nine in its neighborhood. The angle between $\vec{r}$ and $\dot{\vec{r}}$ may change during the flyby with Planet Nine. In fact, they would be perpendicular at the closest point, i.e. at the periapsis, since they are defined relative to Planet Nine. However, we assume that $\vec{r}$ and $\dot{\vec{r}}$ are in the same direction to consider the maximum effect. Then, the ratio of the Newtonian gravity of Planet Nine and the Schwarzschild effect due to Planet Nine ($|\Delta{\ddot{\vec{r}}_{Sch,P9}}|$) is;
\begin{equation}
\frac{|\ddot{\vec{r}}_{P9}|}{|\Delta{\ddot{\vec{r}}_{Sch,P9}}|} \approx 10^{5}.
\label{eq:ratio5}
\end{equation}
In fact, the ratio would be even bigger because of the aforementioned assumption. It does not give a meaningful contribution to the deflection.

\subsection{Geodetic Precession Effect}
Due to the curvature in the spacetime, a vector, for example, the spin vector of a particle orbiting around a massive body will precess, so the direction of the spin vector would not be the same after one period. This is called as geodetic precession effect. This effect also causes the precession on the eccentricity vector or the orbit angular momentum vector of the particle and changes its orbit. This effect is given as~\citep{sosnica};
\begin{equation}
\Delta{\ddot{\vec{r}}_{Geo}} = (1+2\gamma)\left[\dot{\vec{R}}\times\left(\frac{-GM\vec{R}}{c^2R^3}\right)\right]\times\dot{\vec{r}}.
\label{eq:geodetic}
\end{equation}
Here, $\gamma$ is 1, $c$ is the speed of light and $\dot{\vec{r}}$ is the velocity vector of the spacecraft. 

In calculating the geodetic precession effect due to Planet Nine, $GM$ is the gravitational parameter of the Sun, $\vec{R}$ and $\dot{\vec{R}}$ are the position and velocity vectors of the Planet Nine relative to Sun. Since the planar motion is studied, the orbit angular momentum vector of Planet Nine ($\vec{R}\times\dot{\vec{R}}$) is perpendicular to the velocity vector of the spacecraft ($\dot{\vec{r}}$). In the vicinity of Planet Nine, the ratio of Newtonian gravity of Planet Nine and the geodetic precession effect due to Planet Nine ($|\Delta{\ddot{\vec{r}}_{Geo,P9}}|$) is given as follows:
\begin{equation}
\frac{|\ddot{\vec{r}}_{P9}|}{|\Delta{\ddot{\vec{r}}_{Geo,P9}}|} \approx 10^{5}.
\label{eq:ratio6}
\end{equation}
So, the geodetic precession effect due to Planet Nine is negligible. 
\section{Conclusions} 
We studied various effects on a small spacecraft that is sent to gravitationally probe Planet Nine. As a measure of the detection, we used the angular displacement in the trajectory of the spacecraft. First, we looked for the effect of the Sun by analyzing the motion of a spacecraft in the Circular Restricted Three-Body Problem approach for the Sun--Planet Nine-spacecraft system for the two particular initial conditions. We found that the presence of the Sun causes the motion of Planet Nine and this yields larger or smaller angular displacements compared to Planet Nine-spacecraft two-body trajectory depending on the initial condition. However, the angular displacements in both two-body and three-body trajectories are above the detection threshold, so they are detectable. Next, we studied the effects of Kuiper Belt and outer planets, namely Jupiter, Saturn, Uranus, Neptune, as well as non-gravitational perturbations like the drag force, the magnetic force, and the solar radiation pressure. The non-gravitational perturbations depend on the physical properties of the spacecraft and we took those parameters according to a possible technology demonstration mission with solar sail. We found that the effect of outer planets is about one order smaller than the Planet Nine and the effect of Kuiper Belt is negligible. The angular displacement due to the magnetic force is below the detection threshold, so it is not detectable. The leading order angular displacement is due to the solar radiation pressure for the lower spacecraft velocities, and the drag force for the higher spacecraft velocities. In order to reduce the effects of drag and solar radiation pressure and distinguish the effect of Planet Nine, one of the solutions would be to decrease area to mass ratio. Lastly, the General Relativistic effects like the frame-dragging, Schwarzschild, and geodetic precession on the spacecraft trajectory were investigated by comparing their magnitudes with the Newtonian gravity of Planet Nine. We found that the frame-dragging effect has the smallest effect and the Schwarzschild effect due to Sun has the largest effect. However, none of the General Relativistic effects produces a meaningful contribution to the detection.   
\section{Declaration of Competing Interest} 
The authors declare that they have no known competing financial interests or personal relationships that could have appeared to influence the work reported in this paper.
\section{Acknowledgments} 
We would like to thank Tahsin Cagri Sisman, Cetin Senturk and Niyazi Anil Gezer for useful discussions on the topic.




\bibliography{references}{}
\bibliographystyle{elsarticle-num}


%\begin{thebibliography}{00}


%\bibitem{}

%\end{thebibliography}
\end{document}